\documentclass[prb,twocolumn,superscriptaddress, showpacs]{revtex4}

\usepackage{amsmath}
\usepackage{amsfonts}

\usepackage{graphicx}
\usepackage{times}

\newcommand{\be}{\begin{equation} }
\newcommand{\ee}{\end{equation} }
\newcommand{\ba}{\begin{eqnarray} }
\newcommand{\ea}{\end{eqnarray} }

\newcommand{\up}{\uparrow}
\newcommand{\dn}{\downarrow}

\begin{document}


\title{Majorana Zero Modes in 1D Quantum Wires
Without Long-Ranged Superconducting Order}

\author{Lukasz Fidkowski}
\affiliation{Microsoft Research, Station Q, Elings Hall, University of California, Santa Barbara, CA 93106}
\author{Roman M. Lutchyn}
\affiliation{Microsoft Research, Station Q, Elings Hall, University of California, Santa Barbara, CA 93106}
\author{Chetan Nayak}
\affiliation{Microsoft Research, Station Q, Elings Hall, University of California, Santa Barbara, CA 93106}
\affiliation{Department of Physics, University of California, Santa Barbara, CA 93106}
\author{Matthew P. A. Fisher}
\affiliation{Department of Physics, University of California, Santa Barbara, CA 93106}

\date{compiled \today}

\begin{abstract}
We show that long-ranged superconducting order is not
necessary to guarantee the existence of Majorana fermion
zero modes at the ends of a quantum wire. We formulate
a concrete model which applies, for instance, to a semiconducting
quantum wire with strong spin-orbit coupling and Zeeman
splitting coupled to a wire with algebraically-decaying
superconducting fluctuations. We solve this model
by bosonization and show that it supports Majorana fermion
zero modes. We show that electron backscattering in the
superconducting wire, which is caused by potential variations
at the Fermi wavevector, generates quantum phase slips
which cause a splitting of the topological degeneracy
which decays as a power law of the length of the superconducting wire.
The power is proportional to the number of channels in
the superconducting wire. Other perturbations give contributions
to the splitting which decay exponentially with the length of either
the superconducting or semiconducting wires.
We argue that our results are generic and apply to a large
class of models. We discuss the implications
for experiments on spin-orbit coupled nanowires coated with
superconducting film and for LaAlO$_3$/SrTiO$_3$
interfaces.
\end{abstract}

\maketitle


\section{Introduction}

Kitaev \cite{Kitaev01} showed that a class of superconducting
quantum wires supports a pair of Majorana fermion zero modes,
one at each end. Lutchyn {\it et al.} \cite{Lutchyn10}
and Oreg {\it et al.} \cite{Oreg10} discovered that, in the presence of a parallel
magnetic field, semiconducting wires with strong spin-orbit coupling
fall in this class if superconductivity is induced by proximity to
a bulk 3D superconductor, see Fig.~\ref{fig:Bulk-SC}. As a result of the Majorana zero modes,
the ground state is doubly degenerate. The two states differ
by fermion parity, which is not locally measurable; therefore,
they form a protected qubit. Networks of such semiconducting wires
have been proposed for topological quantum information
processing \cite{Alicea11,Sau10b,Hassler10,Bonderson11b}.

Long-ranged superconducting order is an essential feature of these analyses.
Though such order is sufficient, it does not seem necessary.
Protected Majorana zero modes also exist
in models of the $5/2$ fractional quantum Hall state
\cite{Moore91,Nayak96c,Read96,Lee07,Levin07,Bonderson11a}
and in Kitaev's honeycomb lattice spin model \cite{Kitaev06a},
and neither of these systems has long-ranged
superconducting order.
Therefore, one might expect that a quantum wire with strong
superconducting fluctuations but no long-ranged order could
also support Majorana fermion zero modes.
Consider, on the other hand, a spinless one-dimensional Luttinger liquid,
which has algebraic order, i.e. the two-point correlation function
of the superconducting order parameter decays to zero
as a power of the separation, rather than approaching a constant.
Such a system has gapless bulk fermionic excitations,
so if Majorana fermion zero modes were found at the ends
of such a model, there would be nothing
protecting them against small perturbations.
Furthermore, in the absence of superconducting order,
the two states of a pair of Majorana
fermion zero modes would have different electric charges
-- and not merely fermion parity.
Simply changing the electrostatic potential
should cause an energy splitting between states with
different electric charges.
Therefore, one might, instead, conclude that long-ranged
superconducting order is necessary to protect Majorana
fermion zero modes in quantum wires.

In this paper, we show that this is not the case.
We construct a model of a spin-orbit coupled semiconducting wire
in a magnetic field which is coupled to an $s$-wave superconducting
wire with power-law order. The schematic plot of the heterostructure
is depicted in Fig.~\ref{fig:Mustard}. We show that this model
supports Majorana fermion zero modes at the ends of the wire.
However, a single wire does not support a qubit;
at least two wires are needed.
The basic idea is simple. Consider Kitaev's \cite{Kitaev01}
model of a superconducting quantum wire of spinless fermions.
\begin{multline}
\label{eqn:Kitaev+phase}
H =  -t \sum_i (c_{i+1}^\dagger c_i  + c_i^\dagger c_{i+1})\\
        + |\Delta|  \sum_i (e^{i\phi} c_i c_{i+1}  -
        e^{-i\phi} c_i^\dagger c_{i+1}^\dagger)
\end{multline}
Here, $\phi$ is the phase of the superconducting order parameter.
Let us assume, for the moment, that $\phi$ is a constant,
as in Kitaev's original paper \cite{Kitaev01} and in
Refs. \onlinecite{Lutchyn10,Oreg10}.
If we rotate the fermion operators to the local value of the phase
of the order parameter: $c_i \rightarrow e^{i\phi/2}{\tilde c}_i$,
then the Hamiltonian takes the form
\begin{equation}
\label{eqn:rotated-Hamiltonian}
H =  -t \sum_i ({\tilde c}_{i+1}^\dagger {\tilde c}_i  + {\tilde c}_i^\dagger {\tilde c}_{i+1})
        + |\Delta|  \sum_i ( {\tilde c}_i {\tilde c}_{i+1}  -
         {\tilde c}_i^\dagger {\tilde c}_{i+1}^\dagger)
\end{equation}
At the special point $t=|\Delta|$, this Hamiltonian can be
diagonalized by introducing the
Majorana fermion operators $\gamma_{2i-1} =  {\tilde c}_i +  {\tilde c}_i^\dagger$,
$\gamma_{2i} =  ({\tilde c}_i -  {\tilde c}_i^\dagger)/i$:
\begin{equation}
H = i|\Delta|{\sum_i}\gamma_{2i}\gamma_{2i+1}
\end{equation}
These operators satisfy $\gamma_i = \gamma_i^\dagger$
and $\left\{ \gamma_i , \gamma_j \right\} = 2\delta_{ij}$.
Note that $\gamma_1$ and $\gamma_{2N}$ do not appear
in the Hamiltonian. Therefore, the ground state is doubly degenerate:
$i\gamma_1 \gamma_{2N}$ can be either $\pm 1$
while $i\gamma_{2i}\gamma_{2i+1} = -1$ for $1\leq i \leq N-1$.
Apart from the degeneracy of the ground state,
there is a gap $2|\Delta|$ to excitations.
The operators $\gamma_1$ and $\gamma_{2N}$
are Majorana fermion zero modes, and the qubit
which they form, $i\gamma_1 \gamma_{2N} = \pm 1$,
is protected since the two states are distinguished only by fermion
parity, which cannot be measured by a local operation.
Only an operator which acts on both sites $1$ and $N$
can affect it. Away from the special point, $t=|\Delta|$, the
physics is very similar: there is a gap in the bulk
above two nearly degenerate ground states which
have an energy splitting $\sim e^{-Na/\xi}$, where
$\xi$ is inversely proportional to the bulk gap and $a$ is
the lattice spacing. This phase persists to the more
physical $|\Delta|\ll t$ limit.
Electron-electron interactions in the wire
determine the region of the phase diagram
occupied by this phase \cite{Gangadharaiah11,Sela11,Stoudenmire11,Lutchyn11b}.

Now suppose that $\phi$ is a fluctuating dynamical field.
We can still perform a change of variables similar to the
one which we made in going from Eq. \eqref{eqn:Kitaev+phase}
to Eq. \eqref{eqn:rotated-Hamiltonian}. This will remove
the phase of the order parameter from the second term
in Eq. \eqref{eqn:Kitaev+phase}, the pairing term. However,
it will introduce a coupling between the fermions and
gradients of the order parameter.
If these terms can be neglected, then we will have mapped
a model with fluctuating order parameter
to one with fixed order parameter which is decoupled from
the fluctuations of $\phi$; therefore, it will have Majorana fermion zero
modes. However, there are some subtleties involved in
the change of variables from $c_i$ to ${\tilde c}_i$
when $\phi$ fluctuates. These are most easily
handled using a bosonized formulation of the electronic degrees of
freedom in the wire. We find a special point in Sections
\ref{sec:soluble} and \ref{sec:two-wires} at which the bosonized formulation
simplifies and allows us to completely analyze the model.
We then show in Section \ref{sec:perturbations}
that our analysis is qualitatively unchanged by
perturbations which take the system away from the special point.

The technical subtleties alluded to above have a physical origin
related to the conservation of charge.
Note that the ground state
energy has the form
\begin{equation}
\label{eqn:even-odd-energies}
E(N) = N {\cal E} + E_{\rm even, odd} + O(e^{-aL})
\end{equation}
for even and odd electron numbers $N$,
respectively. (See Ref. \onlinecite{Bonderson11c} for
the analogous relation for paired quantum Hall states.)
The signature of Majorana fermion zero modes at the
endpoints of a wire is that $E_{\rm odd}=E_{\rm even}$.
In a superconducting system without zero modes,
we would have $E_{\rm odd}>E_{\rm even}$.
The difference $E_{\rm odd}-E_{\rm even}$
would simply be the energy cost of an unpaired electron.
In the presence of zero modes, this cost vanishes.
As may be seen from (\ref{eqn:even-odd-energies}), however,
{\it a single wire does not have degenerate states}
unless the electrostatic potential is tuned so that ${\cal E}=0$.

If, however, we consider two such wires, then there are two degenerate states
for fixed total electron number without any fine-tuning. Suppose that
there are $2N$ electrons in the system. Let us denote the
energy of the two wires, isolated from each other, by
${E_1}(N)$, ${E_2}(N)$. They are given by (\ref{eqn:even-odd-energies})
with ${\cal E}^{(1)}$, ${\cal E}^{(2)}$ and $E_{\rm even, odd}^{(1)}$,
$E_{\rm even, odd}^{(2)}$ taking the place of ${\cal E}$ and $E_{\rm even, odd}$.
If there are Majorana zero modes at the endpoints
of both wires in isolation, then $E_{\rm odd}^{(1)} = E_{\rm even}^{(1)}$
and $E_{\rm odd}^{(2)} = E_{\rm even}^{(2)}$.
Then
\begin{equation}
\label{eqn:2-wire-degeneracy}
{E_1}(N) + {E_2}(N) = {E_1}(N-m) + {E_2}(N+m)
\end{equation}
for any $m$, so long as ${\cal E}^{(1)} = {\cal E}^{(2)}$.
Now suppose that the two semiconducting wires
are coupled to the same (power-law) $s$-wave
superconducting wire (which is assumed to be much longer
than either semiconducting wire so that it can be coupled
to both while keeping them far apart),
so that the electrochemical potential
must be the same in the two wires.
Then ${\cal E}^{(1)} = {\cal E}^{(2)}$.
Furthermore, Cooper pairs can tunnel
from either semiconducting wire to the superconductor.
Therefore, rather than a degenerate ground state for each value of
$m$ in \eqref{eqn:2-wire-degeneracy}, there will be
two nearly degenerate states, corresponding to an even or odd number
of electrons in each wire. Such a protected qubit exists for
any fixed electron number. If the electron number were
odd, then the two states would correspond, instead, to (a) even electron number
in wire 1, odd in wire 2; and (b) odd electron number
in wire 1, even in wire 2.

These arguments are supported by explicit calculations
in Sections \ref{sec:soluble} and \ref{sec:two-wires}.
First, we show in Section \ref{sec:proximity}
how the topological degeneracy is manifested
when a semiconducting nanowire is coupled to a bulk 3D superconductor.
Pair tunneling between the wire and the 3D superconductor
is represented by a term in the bosonized effective Hamiltonian
of the form
\begin{equation}
H_\text{pair tun.} \propto \sin 2\theta
\end{equation}
where $\theta$ is the bosonic field satisfying $\rho = \frac{1}{\pi}{\partial_t}\theta$,
where $\rho$ is the charge density. The two ground states
of the system correspond to the two minima of $\sin 2\theta$
as a function of $\theta$. As we discuss in Section \ref{sec:proximity},
these two states differ in fermion parity, as expected for
a pair of Majorana zero modes. Furthermore, if the two
ends of the wire are connected to form a ring, then
the ground state degeneracy disappears because
only the equal amplitude superposition of the two minima
is allowed for periodic boundary conditions of the electrons
(while the orthogonal superposition occurs for anti-periodic
electronic boundary conditions). When we turn in Sections
\ref{sec:soluble} and \ref{sec:two-wires} to the case in
which the superconductor is also one-dimensional and, therefore,
does not have long-ranged order, our analysis will
depend on a careful treatment of the target
space of the bosonic fields. The periodicity conditions
satisfied by these fields encode the quantization of charge,
and the ground state degeneracy cannot be counted properly
without accounting for them.
The use of bosonization techniques also requires a careful treatment of locality: putative Majorana modes in a transformed system may simply be a reflection of a spontaneously broken global ${\mathbb Z}_2$ in the original variables, c.f. the duality between the transverse field Ising model and a Majorana wire.
We wish to stress the topological nature of the Majorana degeneracy in our model: no local observable can distinguish the two states.
A key feature of these models is that there is a single-fermion
gap even though there are gapless superconducting phase
fluctuations, as is already apparent in \eqref{eqn:rotated-Hamiltonian}
if the second line is benign (as we show it to be).
This may be viewed as
a form of the ``spin-gap proximity effect'' \cite{Emery97,Emery99}.
This gap protects the Majorana fermion zero modes. However, as we show below, in addition to fermion tunneling events which lift the topological degeneracy
even in models with long-range superconducting order,
there is another error-causing process involving quantum phase slips
which will have a vanishingly small probability of occurring in a bulk 3D superconductor.
The effect of a quantum phase slip in the middle of a superconducting wire can be understood as that of a vortex encircling a pair of Majorana zero modes. Such a process results in reading out the fermionic parity via the Aharonov-Casher effect and effectively leads to a splitting of the degeneracy. We show that backscattering from impurities generates quantum phase slips in the middle of the wire and causes a splitting of the topological degeneracy which decays algebraically with the size of the system rather than exponentially. However, the exponent is proportional to the number of channels in the superconducting wire. Thus, by making a superconducting wire
with sufficiently-many channels, we can make the splitting decay as a high power
of the length.

When this is the case, it is sufficient for the
wires and wire networks of Refs. \cite{Lutchyn10,Oreg10,Alicea11}
to be in proximity to systems with power-law superconducting
order; long-ranged order is not necessary. Consequently,
it may be possible to sputter superconducting grains
onto the semiconducting wire or to coat it with
superconducting film of a finite thickness. This
is important because it may be difficult to tune
a semiconducting wire between topological and non-topological
phases by applying a gate voltage if it is in contact with
a bulk superconductor which will presumably fix its chemical potential.

Recently, it has been shown that quasi-1D wires can be
``written'' on LaAlO$_3$/SrTiO$_3$
interfaces \cite{Cen11} which have substantial
Rashba spin-orbit coupling \cite{Caviglia10}.
These wires show strong superconducting
fluctuations. As we will discuss in detail elsewhere \cite{Fidkowski-in-prep},
a possible model for this system is a spin-orbit coupled quantum
wire in contact with local superconducting regions which fail to percolate
across insulating SrTiO$_3$ but can induce still superconducting
fluctuations in quantum wires at the LAO/STO interface.
Our results imply that these superconducting fluctuations
may be sufficient to support Majorana zero modes
at the ends of such wires if a parallel magnetic field is applied.

\section{A Semiconductor Nanowire Coupled to
a Bulk 3D Superconductor: Bosonized Formulation}
\label{sec:proximity}

\begin{figure}[tb]
\centerline{
\includegraphics[width=2.75in]{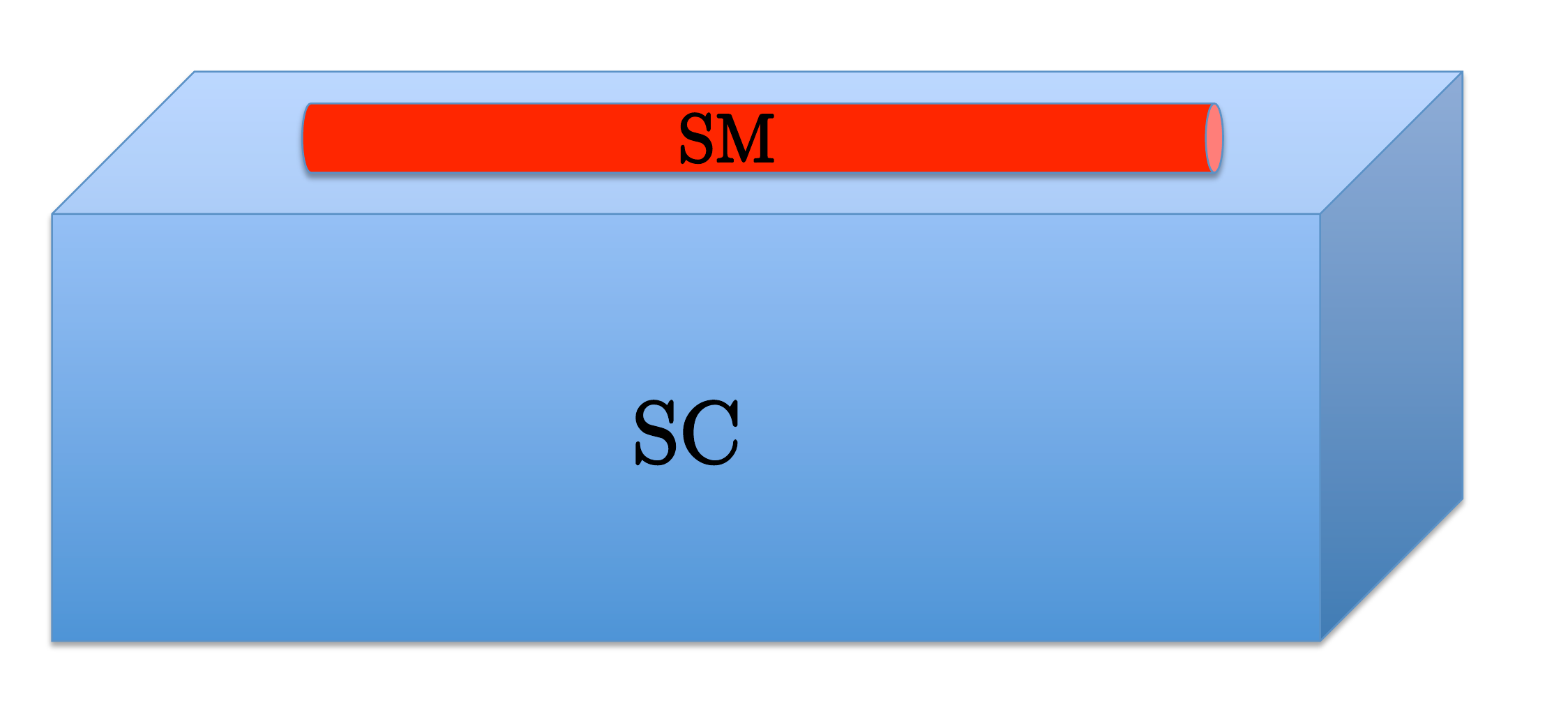}}
\caption{A semiconductor nanowire in contact with a bulk 3D superconductor.}
\label{fig:Bulk-SC}
\end{figure}

Before introducing our model, we briefly review the proposal for realizing Majorana quantum wires in semiconductor-superconductor heterostructures~\cite{Lutchyn10, Oreg10} and recast it in bosonic form. Its basic ingredient is a semiconductor nanowire with strong spin-orbit interactions. Superconductivity is induced via the proximity effect.  The Hamiltonian for the nanowire is ($\hbar=1$):
\begin{align}
\!\!\!H_{\rm NW}\!&=\!\!\!\int_{-L/2}^{L/2} \!d x\, \psi_{\sigma}^\dag(x)\!\!\left(\!-\!\frac{\partial_x^2}{2m^*}\!-\!\mu\!+\!i\alpha \sigma_y \partial_x\!+\!V_x\sigma_x\!\right)_{\!\!\sigma\sigma'}\!\!\!\psi_{\sigma'}(x), \nonumber\\
H_{\rm P}&=\int_{-L/2}^{L/2} dx \left[ \Delta_0 \psi_{\uparrow} \psi_{\downarrow}
+h.c.\right].
\label{eq:H0_prox}
\end{align}
where $m^*$, $\mu$ and $\alpha$ are the effective mass, chemical potential and strength of spin-orbit Rashba interaction, respectively. An in-plane magnetic field $B_x$ leads to a spin splitting  $V_x\!=\!g_{\rm SM}\mu_B B_x/2$, where $g_{\rm SM}$ and $\mu_B$ are the g-factor in the semiconductor and the Bohr magneton, respectively.
In the simplest model for the nanowire, we assume that the semiconductor
nanowire (NW) is in tunneling contact with a bulk 3D superconductor (SC),
as depicted in Figure \ref{fig:Bulk-SC}. Then,
electron tunneling between the NW and the SC leads to the
proximity effect described by the Hamiltonian $H_{\rm P}$.
The superconducting pairing potential $\Delta_0$ is assumed to
be a static classical field and quantum fluctuations of the
superconducting phase are neglected.

The nanowire described by the Hamiltonian $H_{\rm T}=H_{\rm NW}+H_{\rm P}$ can be driven into a non-trivial topological state by adjusting the chemical potential so that it lies in the gap $|\mu|<\sqrt{V_x^2-\Delta_0^2}$. Under these conditions the
Hamiltonian can be projected to the lower band of the two bands which
form as a result of the combined effect of the spin-orbit coupling and magnetic
field. The low-energy limit of this Hamiltonian then takes the same
form as Eq. \eqref{eqn:rotated-Hamiltonian} for low energies $E\ll t$,
assuming $|\Delta|\ll t$ \cite{Lutchyn10}. Therefore,
the topological superconducting phase described by $H_{\rm T}$ harbors Majorana fermion operators $\gamma_L$ and $\gamma_R$
which are zero modes, up to exponential corrections, localized
about the two endpoints:
\begin{equation}
{\gamma_a} = {\gamma_a^\dagger}  \,\, ,
\,\,\, \left\{ \gamma_a , \gamma_b \right\} = 2\delta_{ab}
\end{equation}
\begin{equation}
[{H_T}, \gamma_a ] = 0 + O(e^{-L/\xi}).
\end{equation}
\begin{equation}
\left\{{\gamma_L},{\psi_\sigma}(x)\right\} \sim e^{-|x+L/2|/\xi} \,\, , \,\,\,
\left\{{\gamma_R},{\psi_\sigma}(x)\right\} \sim e^{-|x-L/2|/\xi}
\end{equation}
Here, $\xi$ is the effective coherence length.
The presence of these zero-modes leads to topological degeneracy
up to an exponential splitting energy $\delta E \propto e^{-L/\xi}$.
The two nearly-degenerate states correspond to the two eigenvalues of
$i\gamma_1 \gamma_2$ and have even and odd fermion-parity~\cite{Kitaev01},
respectively, which can be exploited for topological quantum
computation~\cite{Nayak08}.

These results were obtained \cite{Kitaev01,Lutchyn10,Oreg10}
using the properties of the free fermion band structure
embodied by $H_{\rm T}$. We now re-derive them using a bosonic
representation. In later sections, we will use this representation
to analyze the case when there is no long-ranged superconducting order,
unlike in $H_T$. First, we bosonize the semiconductor Hamiltonian \eqref{eq:H0_prox}. In the helical regime corresponding to a large Zeeman gap, $H_{\rm NW}$ can be approximated by projecting the system to the lowest subband and writing the field operator $\Psi(x)\equiv (\psi_{\up}(x),\psi_{\dn}(x))$ as
\begin{align}\label{eq:spinor}
\Psi(x)\!\approx\! \Phi_-(p_F)e^{i p_F x} c_R(x)+ \Phi_-(-p_F)e^{-i p_F x} c_L(x)
\end{align}
where the spinor $\Phi_-(p_F)=\frac{1}{\sqrt 2} \left(-e^{i\kappa(p_F)},1\right)$  and
$\kappa(p_F)~=~\tan^{-1}(\alpha p_F/V_x)$. Substituting \eqref{eq:spinor} into $H_{\rm NW}$, the Hamiltonian can be written in terms of the spinless right and left-moving fermions $c_R(x)$ and $c_L(x)$ and eventually bosonized using $c_{R/L}=\frac{1}{\sqrt{2\pi a}}e^{-i(\pm \phi-\theta)}$:
\begin{align}
H_{\rm NW} &\approx v \int_{-L/2}^{L/2} \!\!\!\!\!\!d x  \left[i c^\dag_L(x)\partial_x c_L(x)\!-\!i c^\dag_R(x)\partial_x c_R(x)\right]\\
&\approx \frac{v}{2\pi} \int_{-L/2}^{L/2} \!\!\!\!\!\!d x  \left[K(\partial_x \theta)^2+K^{-1}(\partial_x \phi)^2\right].
\end{align}
Here $v$ is the fermion velocity $v=p_F(\frac{1}{m^*}-\frac{\alpha^2}{\sqrt{V_x^2+\alpha^2p_F^2}})$ and $K$ the Luttinger parameter for the nanowire.
The fields $\phi$ and $\theta$ satisfy the canonical commutation relation:
\begin{equation}
\label{eqn:commutation}
\left[ \partial_x \phi(x) , \theta(x') \right] = {i\pi} \delta(x-x')
\end{equation}
The charge density and current near wavevector zero
are given by $\rho = \frac{1}{\pi}\partial_x \phi = \frac{1}{\pi}\partial_t \theta$
and $j = -\frac{1}{\pi}\partial_t \phi = \frac{1}{\pi}\partial_x \theta$.
The fields $\phi$ and $\theta$ can be interpreted as the
phase of the density at wavevector
$2k_F$ and the pair field, respectively:
\begin{eqnarray}
\rho_{2k_F}(x) &=&  e^{-2i\phi(x)}\cr
\Psi_{\rm pair}(x) &\equiv& {\psi_\uparrow}(x){\psi_\downarrow}(x) = e^{2i\theta(x)}
\end{eqnarray}

For the Hamiltonian $H_{NW}$, in which electron-electron interactions
in the semiconductor have been neglected, $K=1$, the free-fermion value.
However, the bosonic representation accommodates short-ranged
interactions in the nanowire such as
\begin{equation}
H_{\rm NW \,\,int.} = u \int_{-L/2}^{L/2} \!dx\,
{\psi^\dagger_\sigma}(x) {\psi^{}_\sigma}(x)\,
\psi^\dagger_{\sigma'}(x) \psi^{}_{\sigma'}(x)
\end{equation}
simply by shifting the value of $K$ and rescaling $v$.
$K<1$ for repulsive interactions and $K>1$ for attractive interactions.
The bosonic form for $H_{\rm P}$ in Eq. \eqref{eq:H0_prox} is:
\begin{equation}
H_{\rm P} =  \frac{\Delta_P}{(2\pi a)} \int_{-L/2}^{L/2}\!dx\,
\sin\left(2\theta\right)
\end{equation}
Therefore, $H_{\rm T}$ can be written in the bosonic form
\begin{multline}
\label{eqn:H_T-bosonic}
H_{\rm T} =
 \int_{-L/2}^{L/2} \!\!\!\!\!\!d x  \biggl(\frac{v}{2\pi}\left[K(\partial_x \theta)^2+
 K^{-1}(\partial_x \phi)^2\right] \\ +
\frac{\Delta_P}{(2\pi a)} \sin\left(2\theta\right)\biggr)
\end{multline}

This interaction term, $H_{\rm P}$, is relevant unless there are very strong
repulsive interactions in the nanowire. To be more precise,
the lowest-order RG equation for the dimensionless coupling
$y=2\Delta_P a/v$ is:
\begin{align}
\frac{d y}{d l}=(2-K^{-1})\,y
\label{eqn:lowest-order-RG}
\end{align}
For non-interacting electrons, $K=1$, and even for repulsive interactions
up until $K=1/2$, this is a relevant perturbation. If $y$ is initially
small at short distances, then we can use Eq. \eqref{eqn:lowest-order-RG}
to conclude that $y(l) \sim 1$ at the length scale $l=\ln(\xi/a_0)$,
where the effective coherence length, $\xi$, in the semiconducting nanowire
is given by $\xi \sim a_0 (v/2\Delta_P a_0)^{K/(2K-1)}$. Here, $a_0$ is the
short-distance cutoff, which is the shortest length scale
at which the effective description \eqref{eqn:H_T-bosonic} is
valid. We can take it to be the coherence length or the
Josephson length of the bulk 3D superconductor but, at any rate,
it must be larger than the Fermi wavelength in the semiconducting
wire.

At longer length scales, the field
$\theta$ is pinned to the minimum of $\sin(2 \theta)$.
Since there are two minima, $\theta = -\pi/4, 3\pi/4$,
there are two degenerate ground states in the $L\rightarrow\infty$
limit. These two ground states are related
to each other by the global $\mathbb{Z}_2$ symmetry of
the model, $\theta\rightarrow\theta +\pi$.
To understand this symmetry better, it is helpful to note
that the fermion parity $(-1)^{N_F}$
can be written in the form
\begin{equation}
\label{eqn:fermion-parity}
(-1)^{N_F} = e^{i(\phi(L/2)-\phi(-L/2))}
\end{equation}
Therefore, using the commutation relation \eqref{eqn:commutation},
we see that the fermion parity $(-1)^{N_F}$ generates
the symmetry transformation $\theta\rightarrow\theta +\pi$.
Since the two degenerate ground states corresponding
to $\theta = -\pi/4, 3\pi/4$ are transformed into each other
by fermion parity, the following quantum superpositions
are fermion parity eigenstates:
\begin{equation}
\left| \text{even},\text{odd}\right\rangle = \frac{1}{\sqrt{2}}\Bigl( \left|-\pi/4 \right\rangle
 \pm \left| 3\pi/4\right\rangle\Bigr)
\end{equation}

The ends of the wire are crucial for this qubit.
If we were to connect the two ends of the wire to form
a ring of circumference $L$, then we would expect only a single
ground state, not a degenerate pair. To see that this
is, indeed the case, consider the fermion annihilation
operators:
\begin{equation}
c_{R,L}(x) = \frac{1}{\sqrt{2\pi a}}\, e^{-i(\pm\phi -\theta)}
\end{equation}
Since $\rho=\frac{1}{\pi}{\partial_x}\phi$, the ring will
have even fermion parity if the boundary conditions
on $\phi$ are:
$$
\phi(x+L)=\phi(x)+2n\pi
$$ for integer $n$.
If the fermions have periodic boundary conditions,
$c_{R,L}(x+L)=c_{R,L}(x)$, then the boundary condition
on $\theta$ must be
$$
\theta(x+L)=\theta(x)+2n'\pi
$$
for integer $n'$.
Since constant solutions are allowed for this boundary condition
on $\theta$, the ground state
$\left| \text{even}\right\rangle = \frac{1}{\sqrt{2}}\Bigl( \left|-\pi/4 \right\rangle
 + \left| 3\pi/4\right\rangle\Bigr)$, which is a linear superposition
 of constant solutions, is allowed in this case. This state
 has even fermion parity (\ref{eqn:fermion-parity}),
 so it is consistent with the
 boundary conditions on $\phi$.
If the ring has odd fermion parity, however, then
$\phi(x+L)=\phi(x)+(2n+1)\pi$. Consequently, if 
the fermions have periodic boundary conditions, the boundary
condition on $\theta$ must be $\theta(x+L)=\theta(x)+(2n'+1)\pi$.
This precludes a constant solution. Therefore,
the state $\left|\text{odd}\right\rangle = \frac{1}{\sqrt{2}}( \left|-\pi/4 \right\rangle
 - \left| 3\pi/4\right\rangle)$, which is odd under fermion parity
 (\ref{eqn:fermion-parity}), is not an allowed state
 if the fermions have periodic boundary conditions.
As expected, we conclude that there is only a single ground
state for a ring, in contrast with a line segment which has
a doubly degenerate ground state.

The Majorana fermion zero modes of this system
are manifested on a ring by the presence of a corresponding
state for anti-periodic boundary conditions on the fermions.
If $c_{R,L}(x+L)=-c_{R,L}(x)$, then for odd fermion parity,
$\phi(x+L)=\phi(x)+(2n+1)\pi$, the boundary condition
on $\theta$ must be $\theta(x+L)=\theta(x)+2n'\pi$ for integer $n'$.
This boundary condition allows constant solutions, so
the ground state is $\left|\text{odd}\right\rangle =
\frac{1}{\sqrt{2}}( \left|-\pi/4 \right\rangle - \left| 3\pi/4\right\rangle)$.
Therefore, the ground state with periodic boundary conditions
and the ground state with anti-periodic boundary conditions
have the same energy density and opposite fermion parities.
This can already be seen in the Kitaev chain. On a line segment,
the operators $\gamma_1$ and $\gamma_{2N}$ do not appear in
the Hamiltonian, as we saw in the introduction. On a ring with
periodic boundary conditions, there is a term $it \gamma_{2N} \gamma_1$.
If the boundary conditions are anti-periodic, the term is instead
$-it \gamma_{2N} \gamma_1$. The ground state energy is the same
in both cases, but the ground states differ in fermion parity,
$i \gamma_{2N} \gamma_1 = \pm 1$.

Returning now to the case of open boundary conditions,
we observe that, for finite $L$, these two states are
split in energy because there are instantons which tunnel
between the two minima. The Euclidean action in the strong coupling limit is
\begin{eqnarray} S \!=\! \frac{v}{2\pi}\! \int dx \,d\tau\,\!\left[(\partial_x \theta)^2 \!+\! v^{-2} (\partial_\tau \theta)^2+\frac{y}{\xi^2}\sin(2 \theta)\right]
 \end{eqnarray}
The splitting is then given by $\delta E \propto N_f e^{-S_0}$, where $S_0$ is the action of the Euclidean instanton $\theta_0(x,\tau)$ satisfying $\theta_0(x,-\infty) = -\frac{\pi}{4}$, $\theta_0(x,\infty)=\frac{3\pi}{4}$ and $N_f$ is a prefactor that comes from fluctuations. Clearly the lowest action instanton is translationally invariant, at least away from $x=-L/2, L/2$, so the problem reduces to a $0+1$ dimensional problem, with action
\begin{equation} S_{QM} = \frac{L}{\pi} \int d z \left[ \frac{1}{2} (\partial_{z} \theta)^2 + V(\theta) \right], \end{equation}
where $V(\theta) = \frac{y}{\xi^2} \sin (2 \theta)$ and $z=v \tau$.
Following Ref. \onlinecite{Coleman85},
\begin{equation}
S_0 = \frac{L}{\pi} \int_{-\pi/4}^{3 \pi/4} d\theta \sqrt{2(V(\theta)-E)} = \frac{4\sqrt y}{\pi} \frac{L}{\xi},
\end{equation}
where $E=-y/\xi^2$ is the energy of the minimum of the potential.
The splitting then scales like $\delta E\propto \exp\left(-\frac{4\sqrt y}{\pi} \frac{L}{\xi}\right)$, as expected.

Since $2\theta$ changes by $2\pi$ while the phase of the bulk
superconductor is unchanged, such an instanton
can be interpreted roughly as the motion of a vortex between
the NW and the bulk superconductor. (We say ``roughly''
because our instanton is a spatially uniform phase slip, rather
than a spatially-localized vortex.) Since it causes a transition
between the states $|-\pi/4\rangle$ and $|3\pi/4\rangle$,
it splits the states $|\text{even}\rangle$ and $|\text{odd}\rangle$.
Thus, it can also be interpreted as Majorana fermion tunneling
between the two ends of the wire.

\section{A Single Semiconducting Nanowire Coupled to
an Algebraically-Ordered Superconducting Wire}
\label{sec:soluble}

We now include the effect of quantum fluctuations by replacing the bulk superconductor in the above proposal with an $s$-wave superconducting wire with power-law order. This model preserves the overall $U(1)$ charge symmetry (there is no spontaneous $U(1)$ breaking) and allows for the study of the topological superconducting phase in the particle number-conserving setting. For the sake of concreteness and simplicity,
we will take the Hamiltonian for the superconducting wire to be the attractive-$U$ Hubbard model. However, our results hold for any spin-gapped system
with $s$-wave superconducting fluctuations.

\begin{figure}[tb]
\centerline{
\includegraphics[width=2.75in]{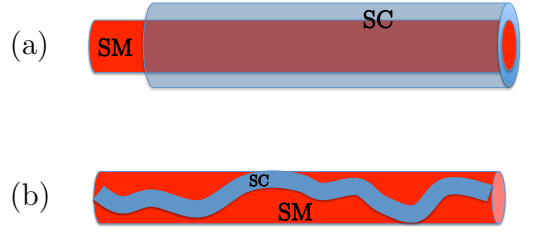}}
\caption{A semiconductor nanowire in contact with a 1D superconducting
wire. The superconducting wire could be a coating which
(a) completely covers the semiconductor or (b) only covers part of
it.}
\label{fig:Mustard}
\end{figure}

We use the standard bosonization procedure for spinful fermions,
with the convention~\cite{Giamarchi_book} that
\begin{equation}
\label{eqn:spinful-bosonization}
\psi_{r,\sigma}=\frac{1}{\sqrt{2\pi a}}e^{-\frac{i}{\sqrt{2}}[(r \phi_{\rho}-\theta_{\rho})+\sigma(r \phi_{\sigma}\!-\!\theta_{\sigma})]}
\end{equation}
where $r=\pm$ and $\sigma=\pm$ for right/left-moving fermion with $\up/\dn$ spin, and $a$ the lattice cutoff. The fields $\phi_{\rho,\sigma}$ and $\theta_{\rho,\sigma}$
satisfy the same commutation relations (\ref{eqn:commutation}).
In terms of these fields, the Hamiltonian for the superconducting
wire can be written as
\begin{align}
H_{\rm SC}&=H^{(\rho)}_{\rm SC}+H^{(\sigma)}_{\rm SC}\label{eq:H_AH_total} \\
H^{(\rho)}_{\rm SC}&=\frac{v_F}{2\pi}\int_{-L/2}^{L/2}dx \left[K_{\rho} (\partial_x \theta_{\rho})^2+ K_{\rho}^{-1}(\partial_x \phi_{\rho})^2\right] \label{eq:H_AH_rho} \\
H^{(\sigma)}_{\rm SC}&=\frac{v_F}{2\pi}\int_{-L/2}^{L/2}dx \left[K_{\sigma} (\partial_x \theta_{\sigma})^2+ K_{\sigma}^{-1}(\partial_x \phi_{\sigma})^2\right] \label{eq:H_AH_sigma} \\
&-\frac{2|U|}{(2\pi a)^2}\int_{-L/2}^{L/2}dx \cos(2\sqrt{2}\phi_{\sigma}) \nonumber
\end{align}
where $v_F$, $a$ and $U$ are the Fermi velocity, the effective cutoff length and the interparticle interaction potential, respectively.

Tunneling between the superconducting wire and
the semiconducting wire can be described using a simple model Hamiltonian
\begin{align}
H_t=t\sum_{\sigma}\int_{-L/2}^{L/2}dx  (\psi^\dag_{\sigma}\eta_{\sigma}+\eta^\dag_{\sigma}\psi_{\sigma}),
\end{align}
where $t$ is the tunneling amplitude and $\psi_{\sigma}$ and $\eta_{\sigma}$ represent fermion annihilation operators in the semiconducting and superconducting systems, respectively. Given that single-electron tunneling into the superconducting wire is suppressed due to the presence of the spin gap $E_{g}$ (see below), the dominant contribution to the action comes from pair hopping. The perturbative expansion in $t$ to second order leads to the following imaginary-time action
\begin{align}\label{eq:pair_hopping1}
S_{\rm PH}\!&=\!- t^2\sum_{\sigma}\!\!\int\!\! dx d \tau dx' d \tau'\! \!\\
&\left[ \psi^\dag_{\sigma}(x,\tau)\psi^\dag_{-\sigma}(x',\tau')\eta_{\sigma}(x,\tau)\eta_{-\sigma}(x',\tau') \!+\! h.c. \right].\nonumber
\end{align}

We now analyze the bosonized action.  First, the spin field $\phi_{\sigma}$
orders as a result of the last term in Eq. \ref{eq:H_AH_sigma}, opening a spin gap $E_g$ in the superconducting wire. The dual field $\theta_{\sigma}$ is disordered, and its correlation function decays exponentially $\langle e^{-\frac{i}{\sqrt 2} \theta_{\sigma}(x,\tau)} e^{\frac{i}{\sqrt 2} \theta_{\sigma}(0,0)} \rangle_{\sigma} \sim a/\sqrt{x^2+(v_F \tau)^2} \exp[-E_g\sqrt{\tau^2+x^2/v_F^2}]$. This allows us to simplify the action \eqref{eq:pair_hopping1} and make a local approximation
\begin{align}\label{eq:pair_hopping}
S_{\rm PH}\!&\approx \!- \frac{\Delta_P}{(2\pi a)} \int d \tau \int_{-L/2}^{L/2}dx \sin\left(\sqrt{2} \theta_{\rho}-2\theta\right)
\end{align}
valid in the long-time limit $|\tau-\tau'| \gg E_g^{-1}$. Here the Cooper
pair hopping amplitude $\Delta_P$ is given by $\Delta_P \sim \frac{t^2}{E_g} \frac{\alpha p_F}{\sqrt{(\alpha p_F)^2+V_x^2}}$ similarly to the proximity-induced gap in the perturbative tunneling limit $t\ll E_g$. If the field $\theta_{\rho}$ were pinned (i.e. $\theta_{\rho}=0$), we would recover the model considered in
Refs.~\onlinecite{Lutchyn10, Oreg10}. In the present case, however,
overall $U(1)$ symmetry is not broken due to the presence of
fluctuating field $\theta_{\rho}$. Henceforth we thus analyze the
following effective low-energy model
\begin{align} \label{LEmodel}
H_{\rm M}&=\frac{v}{2\pi} \int_{-L/2}^{L/2} \!\!\!\!\!\!d x  \left[K(\partial_x \theta)^2+K^{-1}(\partial_x \phi)^2\right]\nonumber\\
&+\frac{v_F}{2\pi}\int_{-L/2}^{L/2}dx \left[K_{\rho} (\partial_x \theta_{\rho})^2+ K_{\rho}^{-1}(\partial_x \phi_{\rho})^2\right]\nonumber\\
&-\frac{\Delta_P}{(2\pi a)} \int_{-L/2}^{L/2}dx \sin\left(\sqrt{2} \theta_{\rho}-2\theta\right),
\end{align}
and study the effect of quantum fluctuations of $\theta_{\rho}$ on the stability of the topological superconducting phase. This model is quadratic, except for
the interaction $\Delta_P$. The dimensionless
coupling $y=2\Delta_P a/v$ has RG equation
\begin{equation}
\frac{dy}{dl} = \left(2 - \mbox{$\frac{1}{2}$}K_\rho^{-1} - K^{-1}\right)\,y
\end{equation}
For $\frac{1}{2}K_\rho^{-1} + K^{-1} > 2$, this interaction is
irrelevant, and we can ignore Cooper pair tunneling between the wires.
However, inter-wire pair tunneling is relevant for
$\frac{1}{2}K_\rho^{-1} + K^{-1} < 2$, which includes
the case of weakly-attractive interactions in the superconducting
wire, $K_\rho \stackrel{<}{\scriptscriptstyle \sim}1$, and weakly-repulsive interactions
in the semiconducting wire, $K \stackrel{>}{\scriptscriptstyle \sim}1$.

This model simplifies significantly at the special point $v_F=v$ and $2K_{\rho}=K$. At this point, one can diagonalize the Hamiltonian (\ref{LEmodel})
by introducing new variables
$\theta_+=\theta_{\rho}/\sqrt 2+\theta$ and $\theta_-=\theta_{\rho}/\sqrt 2-\theta$:
\begin{align}
H&=\frac{v}{2\pi} \int_{-L/2}^{L/2} \!d x  \left[K_{\rho}(\partial_x \theta_+)^2+K_{\rho}^{-1}(\partial_x \phi_+)^2\right]\nonumber\\
&+\frac{v}{2\pi}\int_{-L/2}^{L/2}dx \left[K_{\rho}(\partial_x \theta_{-})^2+ K_{\rho}^{-1}(\partial_x \phi_{-})^2\right]\nonumber\\
&-\frac{\Delta_P}{(2\pi a)} \int_{-L/2}^{L/2}dx \sin\left(2\theta_{-}\right).
\end{align}
The first line of this Hamiltonian describes gapless superconducting
phase fluctuations. The second and third lines, which are
decoupled from these gapless fluctuations, are identical
to the Hamiltonian \eqref{eqn:H_T-bosonic} for the proximity
effect from a bulk 3D superconductor with long-ranged
superconducting order parameter. At this point, the dimensionless
coupling $y=2\Delta_P a/v$ has RG equation
\begin{equation}
\frac{dy}{dl} = (2-K_\rho^{-1})y
\end{equation}
Therefore, a fermionic gap ${\Delta_F} \sim
\frac{v}{a_0}({\Delta_P}{a_0}/v)^{1/(2-K_\rho^{-1})}$
opens up as a result of the coupling between the wires.

\begin{figure}[tb]
\centerline{
\includegraphics[height=1.5in]{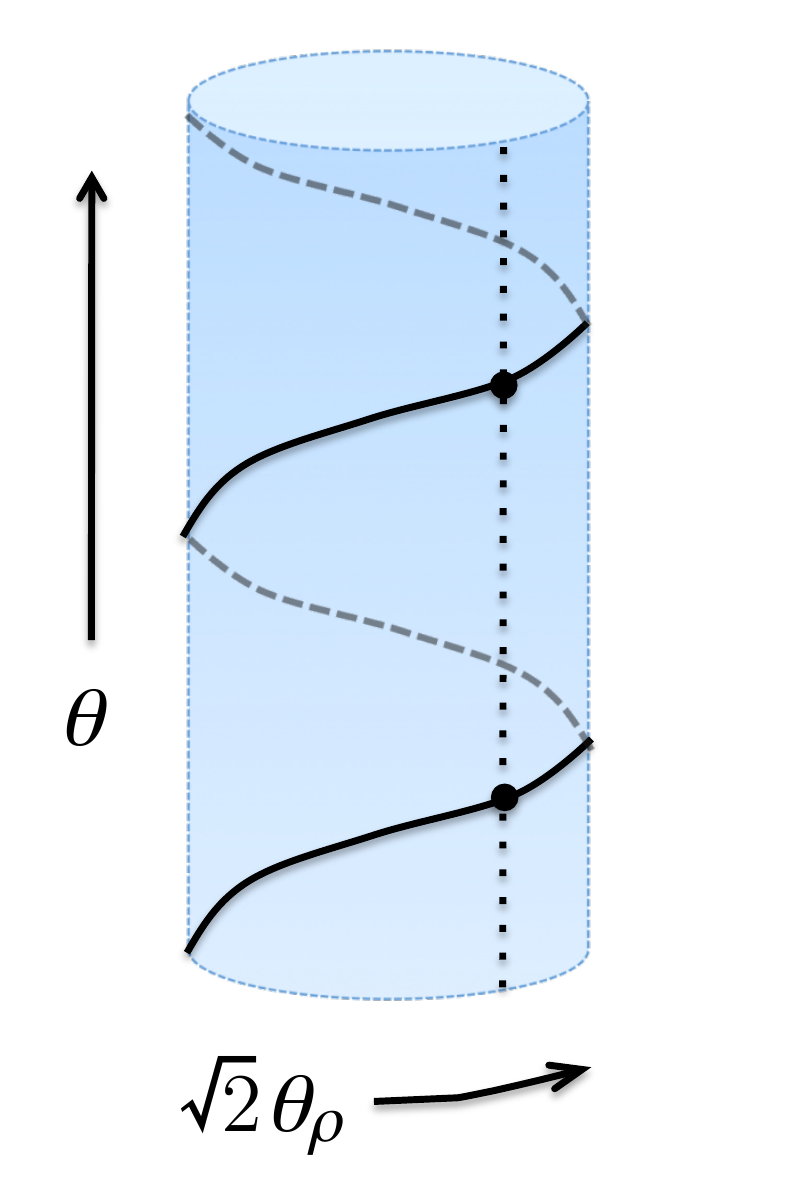}}
\caption{The moduli space of semiclassical vacua is the torus,
which is here depicted as a cylinder with the top and bottom
edges identified. For any fixed $\theta_\rho$, there are
two different semiclassical ground state, depcited by the intersection
points between the dotted vertical line and the line of fixed
$\sqrt{2}{\theta_\rho}-2\theta$ which winds twice
around the cylinder. If $\theta_\rho$ is, indeed,
fixed, as in Section \ref{sec:proximity}, then there is a tunneling
barrier between these two ground states. However, if the total
charge mode $\frac{1}{\sqrt{2}}\theta_\rho + \theta$
can fluctuate, as in Section \ref{sec:soluble},
then the entire line of fixed $\sqrt{2}{\theta_\rho}-2\theta$ is the same quantum
ground state, and there is no degeneracy.}
\label{fig:periodicity}
\end{figure}

The single wire model, however, does not exhibit Majorana
degeneracy without fine tuning of the electrostatic potential.
Semiclassically, this is because the moduli space of low energy field configurations (i.e. those where $\theta_-$ is pinned) has only one connected component.  Naively,
the $\Delta_P \sin(\sqrt{2}\theta_\rho - 2\theta)$ term might lead one to expect two connected components, corresponding to the two minima $\theta = \frac{\theta_\rho}{\sqrt{2}} - \frac{3\pi}{4}, \frac{\theta_\rho}{\sqrt{2}} - \frac{7 \pi}{4}$.  However, these are in fact connected in the $\theta_\rho, \theta$ moduli space, see Fig.\ref{fig:periodicity}.  One can interpolate from one to the other by winding $\sqrt{2} \theta_\rho \rightarrow \sqrt{2} \theta_\rho + 2\pi$ and simultaneously winding $\theta$ half as fast, so that
$\theta \rightarrow \theta + \pi$.  $\theta_-$ remains pinned throughout the interpolation, but the two vacua are exchanged.
Therefore, there is no potential barrier; the field
$\theta_\rho/\sqrt{2} + \theta$ is free to fluctuate along a flat direction
of the potential between these two points.
Consequently, there is just a single vacuum, not two
degenerate states. This reflects the conservation of charge:
when $\theta_\rho/\sqrt{2} + \theta$ has large fluctuations,
the total charge is fixed.

Note that, in the argument above, the two minima were exchanged
if we could identify $\sqrt{2} \theta_\rho \equiv \sqrt{2} \theta_\rho + 2\pi$.
Naively, these two field values are not equivalent
since a shift of $\sqrt{2} \theta_\rho$
by $2\pi$ changes the sign of the fermion according
to Eq. \ref{eqn:spinful-bosonization}. However,
$(\sqrt{2} {\theta_\rho}, \sqrt{2} {\theta_\sigma})\equiv
(\sqrt{2} {\theta_\rho} + 2\pi, \sqrt{2} {\theta_\sigma}+2\pi)$.
Since $\phi_\sigma$ is fixed, $\theta_\sigma$ is disordered,
so there is no energy cost for shifting ${\theta_\sigma}$.
Thus, we can treat $\sqrt{2} {\theta_\rho}$ as $2\pi$ periodic,
rather than $4\pi$ periodic, and the flat direction of the potential
connects the two putative minima.

\section{Two Majorana Wires}
\label{sec:two-wires}

As discussed in the introduction and as we saw in the previous
section, if the electron number is fixed then states with different
electron numbers will not be degenerate without fine-tuning.
However, if we have two semiconducting wires of length $\ell$
coupled to the same superconducting wire, of length $L$,
then there will be two degenerate
states of the system for any fixed total charge. These states
correspond to even or odd electron numbers in each semiconducting
wire, with a constraint that the sum of the parities of the two wires
must equal the parity of the total electron number.
There need not be literally two separate wires. We could instead
have a single wire similar to the spin-orbit coupled
semiconducting wire of Eq. \ref{eq:H0_prox}.
In the regions $-L/2<x<-L/2+\ell$ and $L/2-\ell<x<L/2$,
we would need to adjust the chemical potential so
that $|\mu|<\sqrt{V_x^2-\Delta_0^2}$ and in the
region $-L/2+\ell<x<L/2-\ell$ we would need $|\mu|>\sqrt{V_x^2-\Delta_0^2}$
Then, the system would be in a topological (power-law)
superconducting phase for $L/2-\ell<|x|<L/2-$
and in a non-topological phase for $-L/2+\ell<x<L/2-\ell$.
While we will sometimes call the region $-L/2+\ell<x<L/2-\ell$
the ``non-topological region'', we will usually simply
treat the system as if there were no wire there
the section of non-topological wire in this region has a
qualitatively similar effect to the absence of a wire.

Let us analyze this setup in more detail.
The Hamiltonian for such a system takes the form:
\begin{multline} \label{eq:2wires}
H_{\rm 2 \, wires} =\int_{-L/2}^{-L/2+\ell} dx \biggl(
  \frac{v_1}{2\pi}\left[{K_1} (\partial_x \theta_1)^2 +
K_1^{-1} (\partial_x \phi_1)^2 \right]\\
- \frac{\Delta_{P1}}{(2\pi a)}\,\sin(\sqrt{2}\theta_\rho
- 2{\theta_1})
\biggr)\\
+ \int_{L/2-\ell}^{L/2} dx \biggl(\frac{v_2}{2\pi}\left[{K_2} (\partial_x \theta_1)^2 +
K_2^{-1} (\partial_x \phi_1)^2 \right]\\
- \frac{\Delta_{P2}}{(2\pi a)}\,\sin(\sqrt{2}\theta_\rho
- 2{\theta_2})\biggr)\\
+\int_{-L/2}^{L/2} dx \,
 \frac{v_\rho}{2\pi}\left[{K_\rho} (\partial_x \theta_\rho)^2 +
K_\rho^{-1} (\partial_x \phi_\rho)^2 \right] \\
\end{multline}
The first two lines are the Hamiltonian for the first semiconducting
wire, of length $\ell\ll L$ and its Josephson coupling to the
superconducting wire of length $L$. The third and fourth
lines are the analogous terms for the second semiconducting
wire. The final line reflects the charge degrees of freedom of the Hamiltonian for a wire with power-law superconducting fluctuations. The gapped spin degrees of freedom have been integrated out. In Eq.~\eqref{eq:2wires} we have neglected exponentially small corrections $\propto\exp[-E_g(L-2l)/v_F]$ due to tunneling between the wires, see Sec.~\ref{sec:perturbations} for details.

As in the single wire case, we introduce the fields:
\begin{multline} \label{eq:newfields}
\theta_{+}(x) = \mbox{$\frac{1}{\sqrt{2}}$}{\theta_\rho}(x)
u_{12}(x) + {\theta_\rho}(x)(1-u_{12}(x)) \\ +\,
{\theta_1}(x) {u_1}(x) + {\theta_2}(x) {u_2}(x)\\
\theta_{-}(x) = \mbox{$\frac{1}{\sqrt{2}}$}{\theta_\rho}(x)
u_{12}(x) - {\theta_1}(x) {u_1}(x) - {\theta_2}(x) {u_2}(x)
\end{multline}
where ${u_1}(x)=1$ for $-L/2\leq x\leq -L/2 + \ell$ and ${u_1}(x)=0$
otherwise; ${u_2}(x)=u(-x)$; and $u_{12}(x)={u_1}(x)+{u_2}(x)$.
The field $\theta_{-}(x)$ is only defined for $L/2-\ell \leq |x| \leq L/2$.
Then, for ${v_1}={v_2}={v_\rho}=v$ and
${K_1}={K_2}=2{K_\rho}=K$, the Hamiltonian takes the form:
\begin{multline} \label{eq:H2wires}
H_{\rm 2 \,wires} = \\ \!\int_{-L/2}^{L/2} \!\!\!\!\! dx \biggl(u_{12}(x)
  \frac{v}{2\pi}\left[{K_\rho} (\partial_x \theta_-)^2 +
K_\rho^{-1} (\partial_x \phi_-)^2 \right]\\
+ \mbox{$\frac{1}{(2\pi a)}$}
({\Delta_{P1}}{u_1}(x) + {\Delta_{P2}}{u_2}(x))\,\sin(2{\theta_-})\\
+ \frac{v}{2\pi} \left[{K_\rho} (\partial_x \theta_+)^2 +
K_\rho^{-1} (\partial_x \phi_+)^2 \right]
\biggr)
 \end{multline}
Naively, this Hamiltonian has four semi-classical ground states:
${\theta_-}(x) = {\varphi_1}{u_1}(x) + {\varphi_2}{u_2}(x)$,
with $\varphi_{1,2}=\frac{3\pi}{4},\frac{7\pi}{4}$. However,
by acting with $(-1)^{N_F^{(1)}}$, $(-1)^{N_F^{(2)}}$, and
$(-1)^{N_F^{(1)} + N_F^{(2)}}$
on any one of these states, we can obtain the other three.
Thus, we can form two quantum superpositions of these states
with $(-1)^{N_F^{(1)} + N_F^{(2)}} = 1$ and two with
$(-1)^{N_F^{(1)} + N_F^{(2)}} = -1$.
If we fix the total electron number, then one of these two
sets will be allowed.

The argument which led us to conclude that a single nanowire
has no ground state degeneracy now shows that
the two wire system \eqref{eq:2wires} at fixed electron number has two
(nearly) degenerate ground states. There are two
connected components in the moduli space of low energy field configurations.  Here there is no electrostatic potential breaking the exponentially small degeneracy; the leading contributions instead come from instantons, as we shall see below.   To see this, note first that $\theta_1$ and $\theta_2$ can be pinned to either $\frac{\theta_\rho}{\sqrt{2}} - \frac{3\pi}{4}$ or $\frac{\theta_\rho}{\sqrt{2}} - \frac{7 \pi}{4}$, naively leading to four semiclassical ground states.  However, as above, one can wind $\sqrt{2} \theta_\rho \rightarrow \sqrt{2} \theta_\rho + 2\pi$, thus connecting the ground state $(\varphi_1, \varphi_2) = (\frac{3\pi}{4}, \frac{3\pi}{4})$ with $(\frac{7\pi}{4}, \frac{7\pi}{4})$, and $(\frac{3\pi}{4}, \frac{7\pi}{4})$ with $(\frac{7\pi}{4}, \frac{3\pi}{4})$.  These two equivalence classes cannot be connected to each other, however, since that would require winding $\sqrt{2} \theta_\rho$ by $2\pi$ on only one half of the system, leading to an unwanted monopole in the $\sqrt{2} \theta_\rho$ field.
This monopole can be removed by a phase slip, as we discuss
in Section \ref{sec:impurities}.

Note that there are no bulk operators local in the fermion variables which can distinguish the two nearly degenerate states.  This is because all local terms on, say, wire $1$ must be periodic in $2 \theta_1$; a term that distinguishes the two ground states must necessarily be odd under $\theta_1 \rightarrow \theta_1 + \pi$.  Also, our analysis was for $2$ wires.  A similar analysis for $N$ wires in series,
coupled to the same superconducting wire, produces a degeneracy of $2^{N-1}$.
The ground states correspond to semiclassical vacua
$(\frac{3\pi}{4}+{n_1}\pi,\frac{3\pi}{4}+{n_2}\pi,\ldots,\frac{3\pi}{4}+{n_N}\pi)$,
where $n_i = 0,1$, subject to the condition that the state
$({n_1},{n_2},\ldots,{n_N})\equiv({n_1}+1,{n_2}+1,\ldots,{n_N}+1)$.

As in the proximity-induced case, we can determine the splitting
for two wires via an instanton analysis in a clean system
with only smoothly-varying parameters, which precludes the
possibility of phase slips. According to \eqref{eq:H2wires}, at the soluble point $\theta_+$ decouples, leaving a pinned $\theta_-$ field on each wire.  To tunnel between the two vacua, we need to tunnel from $\theta_- = \frac{3 \pi}{4}$ to $\theta_- = \frac{7\pi}{4}$ along a {\it single} length $\ell$ wire, say wire $1$.  The instanton analysis proceeds exactly as that of the proximity induced case, leading to a splitting $\delta E \sim \exp \left(-\sqrt{K_\rho} \frac{\ell}{\xi} \right)$. In such a process, $2\theta_1$
winds by $\pm 2\pi$ relative to $\sqrt{2}\theta_\rho$. This can be interpreted
as a vortex tunneling between the SM and SC wires, as depicted in
process (a) in Figure \ref{fig:splittings}. It can equivalently can be interpreted
as a fermion tunneling from one end of wire $1$ to the other.

As we discuss in the next section, there is a finite region in parameter space
around the soluble point where a similar analysis applies, resulting again in a splitting exponentially small in $\ell$.

\section{Perturbations About the Soluble Point}
\label{sec:perturbations}

We now show that the Majorana degeneracy is stable against all possible
translationally-invariant perturbations around the soluble point.
We will consider the effect of impurities and phase slips associated
with them in the next section.
The perturbations come in two varieties.  First, there are exponentially small pair hopping terms which involve electrons in different semiconductor nanowires. Secondly, there are couplings between
the semiconducting and superconducting wires which we have not
included in our initial model (\ref{eq:2wires}).
Thirdly, there are shifts of the parameters
which take us away from the point ${K_1}={K_2}=2{K_\rho}$,
${v_1}={v_2}={v_\rho}$. As we will see, the second can be accounted
for with the third.

First, we derive an effective action for interwire pair hopping,
starting with the following microscopic model:
\begin{align}
H_t&=t_1\sum_{\sigma}\int_{-L/2}^{-L/2+\ell} dx_1 (\psi^\dag_{1\sigma}(x_1)\eta_{\sigma}(x_1)+h.c.)\nonumber\\
&+t_2\sum_{\sigma}\int_{L/2-\ell}^{L/2} dx_2 (\psi^\dag_{2\sigma}(x_2)\eta_{\sigma}(x_2)+h.c.).
\end{align}
At second order of perturbation theory in $t$, one obtains cross terms proportional to $t_1t_2$. These exponentially small terms were neglected in Eq.~\eqref{eq:2wires}. Now we take them into account and study their effect on the degeneracy splitting. Consider the term in the Euclidean effective action proportional to $t_1 t_2$:
\begin{align}
S^{(12)}_t&=-2t_1t_2 \int d\tau_1 \int d\tau_2 \int_{-L/2}^{-L/2+\ell} dx_1 \int_{L/2-\ell}^{L/2} dx_2 \nonumber\\
&\times\left(\psi^\dag_{1 \up}(1)\psi^\dag_{2 \dn}(2)\eta_{\dn}(1)\eta_{\dn}(2)+c.c. \right)
\end{align}
Bosonizing the action $S^{(12)}_t$ and integrating out the massive spin fields,
one arrives at
\begin{align}
\!S^{(12)}_t\!&\!=\!\frac{-2t_1t_2 \alpha p_F}{\sqrt{(\alpha p_F)^2+V_x^2}} \! \int d\tau_1\! \int d\tau_2 \!\int_{-L/2}^{-L/2+\ell} dx_1 \! \int_{L/2-\ell}^{L/2} dx_2 \nonumber\\
&\times\cos\!\left(\frac{\phi_{\rho}(1)\!-\!\phi_{\rho}(2)}{\sqrt 2}\!\right)\!\sin \!\left(\!\frac{\theta_{\rho}(1)\!+\!\theta_{\rho}(2)}{\sqrt 2}\!-\!\theta(1)\!-\!\theta(2)\!\right)\nonumber\\
&\times \frac{a}{(2\pi a)^2}\!\frac{\exp\!\left[-E_g\sqrt{(\tau_1\!-\!\tau_2)^2\!+\!\frac{(x_1\!-\!x_2)^2}{v_F^2}}\right]}{\sqrt{(x_1\!-\!x_2)^2\!+\!v_F^2(\tau_1\!-\!\tau_2)^2}}
\end{align}
The dominant contribution to the integral over $\tau_1-\tau_2$ comes from short times ($|\tau_1-\tau_2| \ll (L-2\ell)/v_F$) and can be approximately carried out. The remaining spatial integral is peaked at $x_{1/2}=\pm(L/2-\ell)$, and the action can approximately written as
\begin{multline}
\label{eqn:tunneling1-2}
\!S^{(12)}_t \propto \frac{-t_1t_2}{E_g} e^{-\frac{E_g}{v_F} (L-2\ell)} \frac{\alpha p_F}{\sqrt{(\alpha p_F)^2\!+\!V_x^2}}  \\
\times  \int \! d\tau \cos\!\left(\frac{\phi_{\rho}(-x_0,\tau)\!-\!\phi_{\rho}(x_0,\tau)}{\sqrt 2}\!\right)\\
\times \sin \!\left(\!\frac{\theta_{\rho}(-x_0,\tau)\!+\!\theta_{\rho}(x_0,\tau)}{\sqrt 2}\!-\!\theta(-x_0,\tau)\!-\!\theta(x_0,\tau)\!\right)
\end{multline}
with $x_{0}=L/2-\ell$. In fermionic language, the above expression has a very clear physical interpretation: it corresponds to the Josephson coupling between the ends of the two wires. It is due to single fermion tunneling from one wire to
the other, as depicted in process (b) in Figure \ref{fig:splittings}.
The terms on the second and third lines of (\ref{eqn:tunneling1-2})
are bounded, so such a term causes a splitting
which decays exponentially in $(L-2\ell)$:
$\Delta E \sim e^{-\frac{E_g}{v_F} (L-2\ell)}$.

We now consider interactions between the semiconductor and superconductor wires. We assume that because of the Fermi momenta mismatch in these two systems, one can neglect interactions between the charge and spin-densities at $2k_F^{SC}$
in the superconductor and the corresponding densities at $2k_F^{SM}$ in
the semiconducting nanowire since these interactions will be oscillatory. We will now write down all possible operators couplings between the superconductor and the semiconductor and generate all allowed terms preserving $U(1)$ symmetry. For the superconductor, the charge and spin-density operators are given by
\begin{align}
O_{\rho}&=\psi^\dag_{\up}\psi_{\up}+\psi^\dag_{\dn}\psi_{\dn}=-\frac{\sqrt 2}{\pi }\partial_x \phi_{\rho}\\
O^z_{\sigma}&=\psi^\dag_{\up}\psi_{\up}-\psi^\dag_{\dn}\psi_{\dn}=-\frac{\sqrt 2}{\pi }\partial_x \phi_{\sigma}\\
O^y_{\sigma}\!&\!=\!-i(\psi^\dag_{\up}\psi_{\dn}\!-\!\psi^\dag_{\dn}\psi_{\up})\!=\!\frac{-2}{\pi a}\sin(\sqrt 2 \theta_{\sigma})\cos(\sqrt 2 \phi_{\sigma})\\
O^x_{\sigma}\!&\!=\!\psi^\dag_{\up}\psi_{\dn}\!+\!\psi^\dag_{\dn}\psi_{\up}\!=\!\frac{2}{\pi a}\cos(\sqrt 2 \theta_{\sigma})\cos(\sqrt 2 \phi_{\sigma}),
\end{align}
and the singlet and triplet superconducting pairing operators read
\begin{align}
O_{\rm SS}&=\psi^\dag_{\up}\psi^\dag_{\dn}-\psi^\dag_{\dn}\psi^\dag_{\up}=\frac{1}{\pi a}e^{-i\sqrt 2 \theta_{\rho}}\cos(\sqrt 2 \phi_{\sigma})\\
O^x_{\rm TS}&=\psi^\dag_{\up}\psi^\dag_{\up}+\psi^\dag_{\dn}\psi^\dag_{\dn}=\frac{1}{\pi a}e^{-i\sqrt 2 \theta_{\rho}}\cos(\sqrt 2 \theta_{\sigma})\\
O^y_{\rm TS}&=-i(\psi^\dag_{\up}\psi^\dag_{\up}-\psi^\dag_{\dn}\psi^\dag_{\dn})=\frac{-1}{\pi a}e^{-i\sqrt 2 \theta_{\rho}}\sin(\sqrt 2 \theta_{\sigma})\\
O^z_{\rm TS}&=0,
\end{align}
where SS and TS denote triplet and singlet pairing. We now write down these operators for the semiconductor nanowire. Because of the large Zeeman gap, we perform projection to the lowest subband as explained in Sec.~\ref{sec:proximity}. The
charge- and spin-density operators in the semiconductor now become
\begin{align}
O_{\rho}&=n_R+n_L=-\frac{1}{\pi }\partial_x \phi\\
O^z_{\sigma}&=0\\
O^y_{\sigma}&=\frac{\alpha p_F}{\sqrt{V_x^2+\alpha^2 p_F^2}}(n_R\!-\!n_L)\!=\!\frac{\partial_x \theta}{\pi }\frac{\alpha p_F}{\sqrt{V_x^2+\alpha^2 p_F^2}}\\
O^x_{\sigma}&=\frac{-V_x}{\sqrt{V_x^2+\alpha^2 p_F^2}}(n_R\!+\!n_L)=\frac{\partial_x \phi}{\pi }\frac{V_x}{\sqrt{V_x^2\!+\!\alpha^2 p_F^2}},
\end{align}
and the superconducting pairing operators read
\begin{align}
O_{\rm SS}&=\frac{i\alpha p_F}{\sqrt{V_x^2\!+\!\alpha^2 p_F^2}} c_R^\dag c_L^\dag=\frac{i\alpha p_F }{\sqrt{V_x^2\!+\!\alpha^2 p_F^2}} \frac{e^{-2i\theta}}{\pi a}\\
O^x_{\rm TS}&=O^y_{\rm TS}=O^z_{\rm TS}=0.
\end{align}
The triplet pairing operators vanish because, in our model,
the superconducting wire has a spin gap and, therefore, $\phi_\rho$ is fixed.
Given these operators, one can construct all possible coupling terms between the superconductor and the semiconductor. In addition to the pair-hopping term, which is essential for our proposal to work and was already included in our model
(\ref{LEmodel}) one can have additional couplings which represent various density-density interactions
\begin{align}
H_1&=V_{\rho \rho}\int d x \partial_x \phi_{\rho}\, \partial_x \phi, \label{eq:Hrr} \\
H_2&=V^{(x)}_{\sigma \sigma}\int \frac{d x}{a}\sin(\sqrt 2 \theta_{\sigma})\cos(\sqrt 2 \phi_{\sigma}) \partial_x \phi, \label{eq:Hxx} \\
H_3&=V^{(y)}_{\sigma \sigma}\int \frac{d x}{a}\sin(\sqrt 2 \theta_{\sigma})\cos(\sqrt 2 \phi_{\sigma}) \partial_x \theta. \label{eq:Hyy}
\end{align}
The first term above describes the charge density-density interaction between
the wires whereas the Hamiltonians in the second and third lines correspond to spin-spin interactions. The couplings between current fluctuations
are similar in form to the density-density interactions and have
been have not been included explicitly because their analysis
is so similar. Assuming that $|V^{(x,y)}_{\sigma \sigma}|$ are small compared to $|U|$ in Eq.~\eqref{eq:H_AH_sigma}, the terms \eqref{eq:Hxx} and \eqref{eq:Hyy} can be dropped because the field $\phi_{\sigma}$ orders and $\theta_{\sigma}$ is disordered. Thus, the only coupling that is relevant in the present setup is $H_1$~\eqref{eq:Hrr}. We show below that this quadratic term does not affect the stability of the Majorana modes.

Therefore, a general perturbation is described by the following Euclidean action:

\begin{multline} \label{eq:S2wires}
S^{(E)}_{\rm 2 \, wires} = \int_{-\frac{L}{2}}^{-\frac{L}{2}+\ell} dx \int d\tau \biggl(\frac{K_1}{2\pi v_1} \left[ (\partial_\tau \theta_1)^2 + v_1^2 (\partial_x \theta_1)^2 \right] \\ + \frac{\Delta_{P1}}{2\pi \xi} \sin(\sqrt{2} \theta_\rho-2 \theta_1) + V^{(1)}_{\rho \rho} (\partial_\tau \theta_\rho)(\partial_\tau \theta_1)  \biggr)\\
+\int_{\frac{L}{2}-\ell}^{\frac{L}{2}} dx \int d\tau \biggl(\frac{K_2}{2\pi v_2} \left[ (\partial_\tau \theta_2)^2 +v_2^2 (\partial_x \theta_2)^2 \right]\\+\frac{\Delta_{P2}}{2\pi \xi} \sin(\sqrt{2} \theta_\rho-2 \theta_2) + V^{(2)}_{\rho \rho} (\partial_\tau \theta_\rho)(\partial_\tau \theta_2)  \biggr)\\
+\int_{-\frac{L}{2}}^{\frac{L}{2}} dx \int d\tau \biggl( \frac{K_\rho}{2\pi v_\rho} \left[(\partial_\tau \theta_\rho)^2 + v_\rho^2 (\partial_x \theta_\rho)^2 \right] \biggr) \\
\end{multline}
We re-write this action in terms of the new fields $\theta_+$, $\theta_-$ defined in (\ref{eq:newfields}).  Up to local terms proportional to $\delta(x \pm (\frac{L}{2} - \ell))$, which we drop because they will contribute negligibly to the bulk instanton action, we obtain

\begin{multline} \label{eq:newS}
S^{(E)}_{\rm 2 \, wires} = \int_{-\frac{L}{2}}^{-\frac{L}{2}+\ell} dx \int d\tau \biggl[ A_\tau^{(1)} (\partial_\tau \theta_+)^2 + A_x^{(1)}(\partial_x \theta_+)^2 \\
+ B_\tau^{(1)} (\partial_\tau \theta_-)^2 + A_x^{(1)} (\partial_x \theta_-)^2 + \frac{\Delta_{P1}}{2\pi \xi} \sin(2 \theta_-) \\
+ C_\tau^{(1)} (\partial_\tau \theta_+)(\partial_\tau \theta_-) + C_x^{(1)} (\partial_x \theta_+) (\partial_x \theta_-)   \biggr] \\
+ \int_{\frac{L}{2}-\ell}^{\frac{L}{2}} dx \int d\tau \biggl[ 1\rightarrow 2 \biggr] \\
+\int_{-\frac{L}{2}+\ell}^{\frac{L}{2}-\ell} dx \int d\tau \biggl[ \frac{K_\rho}{2 \pi v_\rho} (\partial_\tau \theta_+)^2+\frac{K_\rho v_\rho}{2 \pi} (\partial_x \theta_+)^2 \biggr]\\
\end{multline}
where $A_\tau^{(1)} = \frac{K_1}{8\pi v_1} + \frac{K_\rho}{4 \pi v_\rho} + \frac{V_{\rho\rho}^{(1)}}{2 \sqrt{2}}$, $A_x^{(1)} = \frac{K_1 v_1}{8 \pi} + \frac{K_\rho v_\rho}{4\pi}$, $B_\tau^{(1)} = \frac{K_1}{8\pi v_1} + \frac{K_\rho}{4 \pi v_\rho} - \frac{V_{\rho\rho}^{(1)}}{2 \sqrt{2}}$, $C_\tau^{(1)} = \frac{K_\rho}{2 \pi v_\rho} - \frac{K_1}{4 \pi v_1}$, $C_x^{(1)} = \frac{K_\rho v_\rho}{2\pi} - \frac{K_1 v_1}{4\pi}$, and similarly with $1$ replaced by $2$.

We see that in the general case $\theta_+$ does not decouple.  However, its action is still quadratic, so we can integrate it out exactly.  We generate the following terms.  On wire $1$ we have

\begin{multline} \label{deltaS1}
\delta S^{(1)} = \int d\omega\, dk \,\frac{2 (C_\tau^{(1)})^2 \omega^4}{A_\tau^{(1)} \omega^2 + A_x^{(1)} k^2} \theta_-^2 \\ + \int d\omega \,dk \,\frac{2 (C_x^{(1)})^2 k^4}{A_\tau^{(1)} \omega^2 + A_x^{(1)} k^2} \theta_-^2
\end{multline}
We obtain an analogous expression for $\delta S^{(2)}$.  We also obtain the following bilinear which couples wires $1$ and $2$:

\begin{multline}
\delta S^{(12)} = \int d\tau_1 \,d\tau_2 \,dx_1 \,dx_2 \\  \biggl[ C_\tau^{(1)} C_\tau^{(2)} \, \left( \partial_{\tau_1} \partial_{\tau_2} \langle \theta_+(1) \theta_+(2) \rangle \right) (\partial_\tau \theta_-(1))(\partial_\tau \theta_-(2)) \\
+C_x^{(1)} C_x^{(2)} \, \left( \partial_{x_1} \partial_{x_2} \langle \theta_+(1) \theta_+(2) \rangle \right)  (\partial_x \theta_-(1))(\partial_x \theta_-(2)) \biggr]
\end{multline}
Here $\langle \theta_+(1) \theta_+(2) \rangle$ is the $\theta_+$ two point function between wires $1$ and $2$.

Suppose now that the coupling $\delta S^{(12)}$ were absent.  Then, as in the instanton analysis of the proximity induced case, we could conclude that the lowest action instanton is translationally invariant on $1$ and $2$ separately; taking $k=0$ in (\ref{deltaS1}) just gives a renormalization of the kinetic term of the $\theta_-^{(1)}$ center of mass mode, and similarly for $\delta S^{(2)}$.  We are interested in an instanton that  tunnels from $\theta_- = \frac{3 \pi}{4}$ to $\theta_- = \frac{7\pi}{4}$ on wire $1$, and remains in the same vacuum on wire $2$.  In the present context, with $\delta S^{(12)}$ absent, such an instanton has the same form as that obtained in the proximity induced case on wire $1$, and is simply constant on wire $2$.  According to that analysis, it leads to a splitting $\delta E \propto \exp (- c \frac{\ell}{\xi})$.

Now put back $\delta S^{(12)}$.  We will show that the change in the instanton (and the change in its action) is of order $\frac{\xi}{L-2\ell}$, and thus negligibly small when $L - 2\ell \gg \xi$.  To de-clutter the following argument, we set all velocities equal to $1$, set all dimensionless constants equal to $1$, and let $\vec{r} = (x,\tau)$.  The action is then

\begin{multline} \label{totS}
S = \int_{\text{1 and 2}} d\tau dx \left[ (\nabla \theta_-)^2+\frac{1}{\xi^2} \sin(2\theta_-) \right] + \delta S^{(1)} + \delta S^{(2)} \\
+ \int d\vec{r}_1 d\vec{r}_2 \biggl[ f(\vec{r}_2 - \vec{r}_1) (\partial_\tau \theta_-(\vec{r}_1)) (\partial_\tau \theta_-(\vec{r}_2)) \\ + g(\vec{r}_2 - \vec{r}_1) (\partial_x \theta_-(\vec{r}_1)) (\partial_\tau \theta_-(\vec{r}_2)) \biggr]
\end{multline}
where
\begin{eqnarray}
f(\vec{r}_2 - \vec{r}_1) &=& \partial_{\tau_1} \partial_{\tau_2} \, \langle \theta_+(\vec{r}_1) \theta_+(\vec{r}_2) \rangle \nonumber \\ g(\vec{r}_2 - \vec{r}_1) &=& \partial_{x_1} \partial_{x_2} \, \langle \theta_+(\vec{r}_1) \theta_+(\vec{r}_2) \rangle
\end{eqnarray}
We do not need the precise forms of $f$ and $g$; rather, all we use is the fact that $|\nabla f(\vec{r}_2 - \vec{r}_1)| < \frac{c'}{(L-2\ell)^3}$ for some constant $c'$, whenever $x_2 - x_1 > L-2\ell$, a condition that is always satisfied in (\ref{totS}), and a similar condition for $g$.  Let us start with the instanton solution discussed above, i.e. the one that minimizes the action with $\delta S^{(12)}$ absent and tunnels between the two vacua only on wire $1$, while staying constant in one of the vacua on wire $2$ .  We plug it into (\ref{totS}) and vary with respect to $\theta_-(\vec{r}_2)$ to obtain

\begin{equation} \label{eom}
\nabla^2 \theta_-(\vec{r}_2) = \frac{2}{\xi^2} \cos (2 \theta_-(\vec{r}_2)) + \frac{\delta S^{(\vec{r}_2)}}{\delta \theta_-(\vec{r}_2)} + h(\vec{r}_2)
\end{equation}
where
\begin{equation}  \label{heq}
h(\vec{r}_2) = \pi \int d\vec{r}_1\, \delta(\tau_1) \partial_\tau f(\vec{r}_2 - \vec{r}_1)
\end{equation}
is sourced by the instanton on $1$, which, because it varies on a time scale $\xi^{-1}$ can be taken to be $\pi \delta(\tau_1)$ for the purposes of this calculation.  From (\ref{heq}) and the previous bound on $|\nabla f|$ we see that $|h(\vec{r}_2)| < \frac{c' \ell}{(L-2\ell)^3}$.  The key point now is that the dimensionful quantity $h(\vec{r}_2)$ is smaller than $\frac{1}{\xi^2}$ by a factor of $\epsilon^2 = \frac{c' \ell \xi^2}{(L-2\ell)^3} \ll 1$.  Thus the inclusion of $h(\vec{r}_2)$ in (\ref{eom}) causes $\theta_-(\vec{r}_2)$ to deviate from its zeroth order solution only by an amount order $\epsilon$.  This is the first step in a perturbative expansion in $\epsilon$ which shows that the inclusion of $\delta S^{(12)}$ causes only a small change, of order $\epsilon$, in the instanton and its action.

Our analysis did not require $2 K_\rho -  K$ to be small
since we were able to integrate out $\theta_+$ exactly
regardless of their values.
Therefore, so long as $\frac{1}{2}K_\rho^{-1} + K^{-1} < 2$,
which implies that $\Delta_P$ is relevant and generates
a coherence length $\xi$, the instanton argument is still valid and
leads to a splitting $\delta E \propto \exp (- c \frac{\ell}{\xi})$.
Thus, the Majorana degeneracy is stable over this
entire region of the phase diagram, which includes
more physically-interesting values than the soluble point.

\section{Electron Backscattering and Phase Slips}
\label{sec:impurities}

We now study the effect of processes in the superconducting wire
which backscatter a right-moving electron into a left-moving
one or vice versa. We can include the effect of an electrostatic potential
in the superconducting wire by adding a term to the
action:
\begin{multline}
\label{eqn:electrostatic-potential}
H_{\rm pot} = \int \! dx\
V(x)\, {\psi^\dagger_\sigma}(x){\psi_\sigma}(x)\\
= \int\!dx\,V(x)\,\left[ {\psi^\dagger_{R\sigma}}(x)\psi_{R\sigma}(x)
 + {\psi^\dagger_{L\sigma}}(x)\psi_{L\sigma}(x) \right.\\ + \left.
e^{-2i{k_F}x} {\psi^\dagger_{R\sigma}}(x)\psi_{L\sigma}(x)
+ e^{2i{k_F}x} {\psi^\dagger_{L\sigma}}(x)\psi_{R\sigma}(x) \right]\\
= \int\! dx\,V(x)\left[\mbox{$\frac{\sqrt{2}}{\pi}$}\partial_x \phi_\rho
+ 2\cos\bigl(\sqrt{2}\phi_\rho+2{k_F}x\bigr)\cos\!\sqrt{2}{\phi_\sigma}\right]\\
=  \int\! dx\,V(x)\left[\mbox{$\frac{\sqrt{2}}{\pi}$}\partial_x \phi_\rho
+ 2\cos\bigl(\sqrt{2}\phi_\rho+2{k_F}x\bigr)\right]
\end{multline}
In going from the penultimate line to the final one, we have used the
fact that there is a spin gap in the SC wire which pins the value of ${\phi_\sigma}$.
The first term in the final line is harmless and can be absorbed by shifting
$\phi_\rho$ which corresponds to a shift of the chemical potential.
Therefore, we will ignore this term from now on.
The second term in the final line causes $2\pi$ phase slips
in the order parameter in the superconducting wire, $e^{i\sqrt{2}\theta_\rho}$,
since
\begin{equation}
\left[\sqrt{2}{\phi_\rho}(x), \,{\partial_x}\!\left(\sqrt{2}{\theta_\rho}(x')\right)\right]
= -2\pi i \delta(x-x')
\end{equation}
This equation expresses the fact that when an electron in a 1D system
is backscattered, a $2\pi$ phase slip occurs.

\begin{figure}[tb]
\centerline{
\includegraphics[width=3.25in]{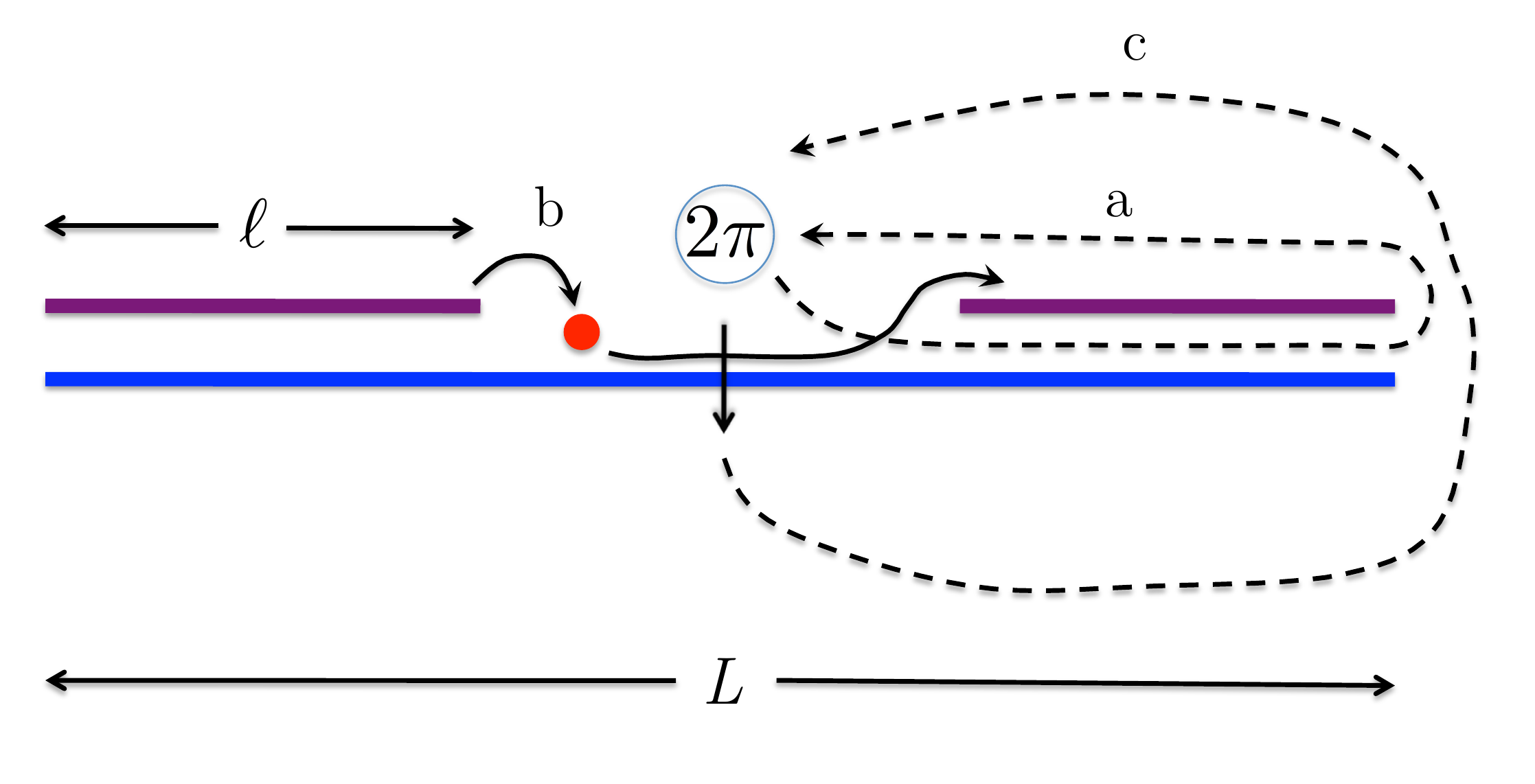}}
\caption{A schematic depiction of the different processes which split the two states of the qubit. (a) A vortex can tunnel between a semiconducting wire and the superconducting wire.
This causes a splitting which is exponentially-small in $\ell$.
(b) An electron can tunnel from one semiconducting wire to another through the superconducting wire. This causes a splitting which is
exponentially-small in $L-2\ell$. (c) A vortex can tunnel through
the superconducting wire between the semiconducting wires as
a result of an electron backscattering process.
This causes a splitting which decays as a power of $L$.
This model applies also to a situation in which there is
a single semiconducting wire which extends from $-L/2$ to $L/2$
and is in a topological phase for $L/2 - \ell < |x| < L$ (similar to
our two SM wires) and is
in a non-topological phase for $|x| < L/2 - \ell$ (similar to
our superconducting wire).}
\label{fig:splittings}
\end{figure}

These phase slips cause transitions between the two states
of the qubit (or, in the fermion parity basis, they cause a splitting
between the two states). At a technical level, this
occurs because a phase slip at the origin causes
$\sqrt{2}{\theta_\rho}$ to wind by $2\pi$ on half of the system.
Then $\theta_1$ can wind by $\pi$ (while remaining at
the minimum of the cosine potential), and the system will make
a transition from the state
$(\frac{3\pi}{4}, \frac{3\pi}{4})\equiv(\frac{7\pi}{4}, \frac{7\pi}{4})$ to
$(\frac{3\pi}{4}, \frac{7\pi}{4})\equiv(\frac{7\pi}{4}, \frac{3\pi}{4})$.
At a more physical level, when a phase slip occurs, a vortex
tunnels across the wire quantum-mechanically. Since there is
no barrier for a vortex to move outside the wire, a vortex
which tunnels through the midpoint of the SC
wire can then encircle half of the SC wire, along
with the NW which is in contact with that half
of the SC wire. The vortex thereby measures the fermion
parity of that NW by the Aharonov-Casher effect,
as is depicted schematically in process (c) in
Figure \ref{fig:splittings}.

Note that if the phase slip occurs between $-L+\ell$
and $L-\ell$ where the semiconducting wire is non-topological
or is absent, there will only be gradient energy in
$\theta_\rho$ (or its dual equivalent, fluctuation energy
in $\phi_\rho$). However, if the phase slip occurs at a point
$x$ satisfying $|L-\ell|<|x|<|L|$, where there is a topological
region of wire, then it will put a kink
in $\sqrt{2}\theta_\rho$ in a region where it
is locked by the potential $\sin(\sqrt{2}{\theta_\rho}-2{\theta_1})$.
Due to the energy cost of a kink, this will leave the system
in a higher energy state. The kink is simply a fermion excited
above the gap. In order to return to a ground state, another
instanton or an anti-instanton must occur. However, this double
process does not mix or split ground states.

We will consider three different types of potentials $V(x)$
which can backscatter electrons. First, we consider a single
impurity. For simplicity, we will focus on the case of a
$\delta$-function impurity at the origin, $V(x)=\frac{v}{2}\,\delta(x)$,
but the physics will be the same for
any potential which is non-zero only in a region
of length much less than $L-2\ell$ near the middle of
the SC wire. Then the Hamiltonian (\ref{eqn:electrostatic-potential})
takes the form:
\begin{equation}
H_{\rm 1-imp.} =
 v\cos\bigl(\sqrt{2}\phi_\rho(0)\bigr)
\end{equation}
The RG equation for $v$ follows from the scaling dimension
for $\cos(\sqrt{2}\phi_\rho)$:
\begin{equation}
\frac{dv}{dl} = \left(1-\mbox{$\frac{1}{2}$}{K_\rho}\right)v
\end{equation}
For ${K_\rho}>2$, this is irrelevant; in the large-$L$, low-temperature
limit the superconductor heals itself and the
backscattering amplitude goes asymptotically to zero.
However, for ${K_\rho}<2$, the SC wire is effectively broken
in two by the impurity. The qubit is then lost. Therefore, it is necessary
to have sufficiently strong attractive interactions in the SC wire
that ${K_\rho}>2$. Even when this is satisfied, the backscattering
amplitude vanishes as a power-law in the system size, not
exponentially. Since backsattering/phase slip processes
cause transitions between the two different states of the qubit
in the phase basis, they causes an energy splitting between
states of different fermion parity:
\begin{equation}
\label{eqn:impurity-splitting}
\Delta E \propto \bigl\langle v\cos\bigl(\sqrt{2}\phi_\rho\bigr) \bigr\rangle
\propto \frac{|v|}{L^{{K_\rho}/2}}
\end{equation}
Since $\phi_\rho$ is fixed at the ends of the SC wire (since no current
flows off the ends), the one-point function for $\cos\sqrt{2}\phi_\rho$
has the $L$ dependence show above.

Now suppose, instead, that there is a random distribution
of impurities so that
\begin{equation}
\overline{V(x) V(x')} = W \delta(x-x')
\end{equation}
Then, we replicate the action by introducing an additional
index $\alpha$ on the field $\phi_\rho^\alpha$ with $\alpha=1,2,\ldots,N$.
We will take $N\rightarrow 0$ at the end of the calculation
in order to take the quenched average over all realizations of the disorder.
The disorder-averaged effective action takes the form:
\begin{equation}
S_{\rm random} =
 \int\! d\tau\,d\tau' dx\, W\cos\bigl(\sqrt{2}\!\left[\phi^\alpha_\rho(x,\tau)
 -\phi^\beta_\rho(x,\tau')\right]\bigr)
\end{equation}
The RG equation for $W$ is:
\begin{equation}
\frac{dW}{dl} = \left(3-{K_\rho}\right)W
\end{equation}
Thus, we need a larger $K_\rho$ for the superconductivity
to survive a random distribution of impurities, and if
${K_\rho} > 3$ is satisfied, then there will be an energy splitting:
\begin{equation}
\label{eqn:random-splitting}
\Delta E \propto  \frac{W}{L^{{K_\rho}-2}}
\end{equation}

Thus far, we have focused on backscattering by impurities,
which effectively create weak spots in the wire where a vortex
can tunnel through. However, even in a completely clean system,
there is some amplitude for backscattering. For instance, let us suppose
that $V(x)$ is constant near the middle of the wire and goes to
zero smoothly near the ends. To make this concrete, let us
take $V(x)=V_0$ for $|x|\ll L/2$ and $V(x)=0$ for $|x|=L/2$.
We will assume that $V(x)$ varies smoothly, so that
the Fourier transform ${\tilde V}(q) \sim e^{-{q^2}b^2}$
for $q\gg 1/L$, where $b>\xi$. Then, from the last line
of Eq. \ref{eqn:electrostatic-potential}, we expect a splitting
\begin{equation}
\Delta E \propto  \int_{-\frac{L}{2}+\ell}^{\frac{L}{2}-\ell}\!dx\,
\frac{V(x) \cos 2{k_F}x}{\left(\frac{L}{2}-|x|\right)^{{\frac{K_\rho}{2}}-2}}
\,\, < \,\, \frac{e^{-4{k_F^2}b^2}}{\ell^{{\frac{K_\rho}{2}}-2}}
\end{equation}
Therefore, as the potential becomes smoother and smoother,
the splitting which it induces through electron backscattering/phase slips
goes exponentially to zero with the length scale $b$ over which
the potential varies. Inhomogeneities enhance backscattering,
as we saw in Eqs. \ref{eqn:impurity-splitting}, \ref{eqn:random-splitting}.

\section{Discussion}
\label{sec:discussion}

As we saw in the Section \ref{sec:impurities}, the effects of
electron backscattering by impurities can be mitigated by
making $K_\rho$ large. In a superconducting wire,
$K_{\rho}=2\pi \sqrt{A_w \rho_s \kappa} \propto {k_F^2}{A_w}\propto N_{\rm channels}$ and $v=\sqrt{A_w \rho_s/ \kappa}$ with
$A_w$,  $\rho_s$ and  $\kappa$ being the cross-sectional area,  superconducting stiffness and compressibility, respectively\cite{Zaikin97}. Therefore,
if the superconducting wire has enough channels or,
equivalently, if the superconducting wire has sufficiently large cross-sectional area
and/or sufficiently large superfluid density, we can have
$K_\rho$ large. Although this is not as good as exponential
decay as a matter of principle, it may be just as good as a practical
matter. This may be important since it could be very difficult to
tune the chemical potential appreciably in the semiconducting wire
(which is necessary to move the Majorana zero modes)
if it is in contact with a bulk 3D superconductor.
Furthermore, coating the semiconducting wires
with superconducting material, as depicted in Figure \ref{fig:Mustard},
may be the easiest way to make a complex
network of wires (especially a three-dimensional network)
which is in contact with a superconductor
\footnote{We thank Charles Marcus for his colorful culinary metaphor
comparing the situation in Fig. \ref{fig:Mustard}b
to mustard on a hot dog.}. However, such
an architecture will necessarily be, at best, an algebraically-ordered
superconductor (except, perhaps, at the lowest temperatures,
at which the coupling between wires causes a crossover to
3D superconductivity). Therefore, it is significant that our results
show that such a network supports Majorana fermion nearly-zero modes
and that their splitting can be made small (albeit not exponentially so).

We also note that it is only important that $K_\rho$
be large in the regions between the topological
semiconducting wire segments. In the topological
semiconducting wire segments, the phase is locked so
that $2\pi$ phase slips cannot occur (although harmless
$4\pi$ phase slips can occur). Therefore, one
can imagine a scenario in which the topological
segments are coated with a thin superconducting film
while the non-topological segments between them
are in contact with essentially bulk 3D superconductors.
This would lead to a protected topological qubit, although
it would be difficult (if not impossible) to move the Majorana
zero modes since that would involve tuning the chemical potential in
the non-topological regions -- which are in contact
with bulk 3D superconductors -- to drive them into the topological phase.
One may alternatively, in a system in which a semiconducting
wire is coated with a thin layer of superconducting material,
use a gate voltage to occupy a large (even) number of sub-bands
of the semiconducting wire in the non-topological regions.
This would lead to a large effective $K_\rho$ for the combined
superconductor-semiconductor system in the non-topological
regions and, therefore, a large power for the decay of the
splitting due to phase slips in these regions.

As the previous sentence anticipates,
our methods should be generalizable to multi-channel
{\it semiconducting} wires \cite{Lutchyn11a,Potter10,Potter11, Wimmer_PRL10}.
They should also apply to a semiconducting wire which
is near a superconducting grain (as in a model of quasi-1D wires in
LAO/STO interfaces\cite{Fidkowski-in-prep}). If the linear size of the grain,
$r$, is smaller than the superconducting coherence length, $\xi$,
then we can treat the grain as a zero-dimensional system.
Suppose that the wire also has length $r$. Then the Hamiltonian
for the wire coupled to the grain is simply \eqref{eqn:Kitaev+phase}
with $\phi$ independent of position $i$ but dependent on
time. There will also be a charging energy $U(N-{N_0})^2$
which causes $\phi$ to fluctuate. There will be no long-ranged order
in the superconducting grain, but it can still induce a single-fermion
gap in the semiconducting wire. Of course, if the wire has length
$L\gg\xi$, then the grain will only change the behavior of a short
section of the wire, and the two ends of this section will be relatively
close to each other. But if the wire passes near many such grains,
then they can induce a single-fermion gap in the wire. If the
coupling between the grains is large compared to their charging energies,
then, in the long-wavelength limit, the grains will develop
algebraic order. The superconducting grains can be
modeled by a superconducting wire, and this situation can be modeled
with the Hamiltonian of Section \ref{sec:soluble},
but with a very small velocity \cite{grains-wires}. If there is ohmic dissipation,
then the grains may not even have power-law superconducting order
but may have exponentially-decaying superconducting correlations.

In fact, we will have Majorana zero modes in a system with
exponentially-decaying superconducting correlations if
we simply take our model to finite temperature. Then,
the $\theta_+$ field in Eq. \eqref{eq:H2wires}
will have exponentially-decaying correlations, with
a correlation length inversely proportional to the temperature.
The $\theta_-$ field will still be pinned to a minimum of the potential,
but it will be possible for the system to be thermally-excited over
the barrier from one minimum to the other. Therefore, if
$\Delta_F$ is the bulk single-fermion gap,
there will be a contribution due to processes (a) and (b)
in Fig. \ref{fig:splittings} to
the coherence time for a Majorana qubit of order $\sim e^{-{\Delta_F}/T}$,
just as if there were long-ranged superconducting order.
However, there will also be a contribution from quantum
phase slips, process (c), which will increase with temperature
as $T^{K\rho/2}$ for a single impurity and $T^{K\rho-2}$
for a random distribution of impurities.
We similarly expect Majorana fermion zero modes to
survive in two-dimensional
structures in which a superconducting gap is induced via the proximity
effect to stabilize a phase with Ising anyons
\cite{Fu08,Sato09,Sato10,Sau10a, Alicea_PRB10,Qi_PRB10}
but long-ranged superconducting
order is disordered by quantum or thermal fluctuations.
If the single-particle gap remains, then the Majorana fermion
zero modes associated with the Ising anyons could survive.
However, quantum phase slips are suppressed \cite{Hermele05}
and, therefore, the splitting will be exponentially, rather than
algebraically, decaying.
Of course, there is nothing surprising about having protected
Majorana zero modes in a system with no long-range order
or even algebraic order, since this is precisely the case
with any true topological phase of matter, as in the examples
mentioned in the introduction. However, the particular route
which we have found to such a system is new and interesting.

In this paper, we have shown that a gapless system can be nearly as
good as a fully-gapped one at supporting protected
Majorana fermion zero modes. It is an interesting open question
whether a gapless system might be capable
of supporting protected degrees of freedom which
cannot occur in fully-gapped 1D systems \cite{Fidkowski11,Turner11,Chen11}.

{\it Note added:} After the initial version of this paper
appeared on the arXiv, several other papers \cite{Cheng11,Tsvelik11,Sau11}
on related topics were submitted to the arXiv.

\acknowledgements
We would like to thank Leon Balents, Meng Cheng, Paul Fendley,
Alexei Kitaev, Steve Kivelson, and Jeremy Levy for discussions.
C.N. is supported by the DARPA QuEST program
and the AFOSR under grant FA9550-10-1-0524.
M.P.A.F. is supported by NSF grant DMR-0529399.

\bibliography{Majorana}
\end{document}